\documentclass[12pt]{article}
\usepackage{fullpage}
\usepackage{amsmath}
\usepackage{amssymb}
\usepackage[dvips]{epsfig}

\def\##1{\underline{#1}}
\def\=#1{\underline{\underline{#1}}}

\def\+#1{\underline{\bf #1}}
\def\*#1{\underline{\underline{\bf #1}}}

\def\r#1{(\ref{#1})}
\def\l#1{\label{#1}}
\def\c#1{\cite{#1}}

\def\le{\left(}
\def\ri{\right)}
\def\les{\left[}
\def\ris{\right]}
\def\lec{\left\{}
\def\ric{\right\}}

\def\.{\mbox{ \tiny{$^\bullet$} }}

\def\epso{\epsilon_{\scriptscriptstyle 0}}

\def\muo{\mu_{\scriptscriptstyle 0}}

\def\curl{\nabla\times}

\def\curl{\nabla\times}

\def\gmet{g_{\alpha\beta}}
\def\gmetm{{}^{(m)}\tilde{g}_{\alpha\beta}}
\def\fmetm{{}^{(m)}d_{\alpha\beta}}
\def\tr{(ct,\#r)}
\def\ok{(\omega/c,\#k)}

\def\calX{{\cal X}}
\def\calXm{{}^{(m)}{\cal X}}

\def\gammam{{}^{(m)}\={\tilde\gamma}}
\def\Gammam{{}^{(m)}\#{\tilde\Gamma}}

\def\mm{\,{}^{(m)}}

\begin{document}

\begin{center}

{\bf {\Large  Gravitation and Electromagnetic Wave Propagation with Negative Phase Velocity}}

 \vspace{10mm} \large

Tom G. Mackay\footnote{Corresponding Author. Fax: + 44 131
650 6553; e--mail: T.Mackay@ed.ac.uk.}\\
{\em School of Mathematics,
University of Edinburgh, Edinburgh EH9 3JZ, UK}\\
\bigskip
 Akhlesh  Lakhtakia\footnote{Fax: +1 814 863 4319; e--mail: akhlesh@psu.edu; also
 affiliated with Department of Physics, Imperial College, London SW7 2 BZ, UK}\\
 {\em CATMAS~---~Computational \& Theoretical
Materials Sciences Group\\ Department of Engineering Science and
Mechanics\\ Pennsylvania State University, University Park, PA
16802--6812, USA}\\
\bigskip
Sandi Setiawan\footnote{Fax: + 44 131
650 6553; e--mail: S.Setiawan@ed.ac.uk.}\\
{\em School of Mathematics,
University of Edinburgh, Edinburgh EH9 3JZ, UK}\\

\end{center}

\vspace{4mm}

\normalsize

\begin{abstract}

Gravitation  has interesting consequences for electromagnetic wave
propagation in a vacuum. The propagation of plane waves with phase
velocity directed opposite to the time--averaged Poynting
vector is investigated for a generally
 curved spacetime.
Conditions for such negative--phase--velocity (NPV) propagation are
established in terms of the spacetime metric components for general and special cases.
 The negative
energy density implications of NPV propagation are discussed.

\end{abstract}


\normalsize \vspace{4mm}


\vspace{4mm}

\bigskip
\section{Introduction}
Negative refraction of a plane wave at the planar interface of two linear, isotropic,
 homogeneous materials is said to occur when the projections of the real parts of the wave vectors
of the incident and the refracted plane waves on the interface normal are
 oppositely directed. Then the real part of the wave vector and
the time--averaged Poynting vector are parallel in one material, but antiparallel in the other
 \c{LMW}. We call the latter kinds as
{\em negative--phase--velocity (NPV) materials\/}, but at least two other names have
common currency too:
left--handed materials, and
negative--index materials. In order to extend the phenomenon of negative refraction to anisotropic materials, NPV materials are characterized
by the negativity of the projection of the real part of the wave vector on the time--averaged Poynting vector.

Since the beginning of the year 2000 \c{SSS}, NPV materials have excited much theoretical as well as experimental interest. Initial disbelief
and criticism  in some sections of the electromagnetics research
community \c{Garcia,Valanju} eventually gave way to widespread, but perhaps still not universal, acceptance
of NPV materials with unequivocal demonstrations by several independent groups \c{NPV_expt1}--\c{PS04}. A simplistic expression
of the (monochromatic) electromagnetic energy density turns out to yield negative values \c{Ziol}, which are generally held as impossible in the electromagnetics
research community, but more sophisticated
investigations  indicate that  the electromagnetic energy
density in NPV materials is indeed positive when account is taken of the  frequency--dependent constitutive properties \c{Rupp}.

Perhaps the potentially most useful application of NPV materials is for the
so--called perfect lenses \c{Pen}. Once satisfactorily designed
and fabricated, such lenses~--~although not really perfect \c{Lpl,WYWN}~---~ could find widespread use
in modern optics, for communications, entertainment, and data storage and retrieval.
More uses would emerge with ongoing research on anisotropic NPV materials, particularly with negligibly small dissipation
in certain frequency ranges.

Instead of concentrating on devices,
 we turned our attention to the marriage
of the special and the general theories of relativity (STR and GTR) and NPV propagation
of electromagnetic fields. We found, a few months ago,
that materials that appear to be of the non-NPV type to
relatively stationary  observers  can appear to be of the NPV type to observers
moving with uniform velocity \c{ML04a}.
That result permitted us to
envisage STR negative
refraction being exploited  in astronomical scenarios \c{ML04b} such as for
the remote sensing of planetary and asteroidal surfaces from space stations.
Application to remotely guided, extraterrestrial mining and
manufacturing industries can also be envisioned. Furthermore, many
unusual astronomical phenomenons would be discovered and/or explained
via  STR negative refraction to
interpret data
collected via telescopes.

Ordinary vacuum (i.e., matter--free space) appears the same to all observers
moving at constant relative velocities. Therefore, NPV propagation in vacuum
cannot be observed by such observers.  This could lead one to believe that NPV propagation is impossible
in huge expanses of interstellar space. However, gravitational fields from nearby massive objects will
certainly distort electromagnetic propagation, which is a principal tenet of the GTR and is indeed used nowadays in GPS systems,
so that NPV propagation under the influence of a gravitational field required investigation. In a short
communication \c{JPA1}, we derived a condition
for NPV propagation to occur along a specific
direction in a region of spacetime, with the assumption of a piecewise uniform but otherwise general spacetime metric. As the
consequences of such a possibility are highly relevant to further exploration of outer space as well as for industrial operations therein,
we undertook a more general study, the results of which are being reported here.

The plan of this paper is as follows: In Section 2,
electromagnetism in generally curved spacetime is transformed from
a covariant to a noncovariant formalism, wherein vacuum resembles
a bianisotropic ``medium" which enables planewave propagation to
be examined using standard techniques. A piecewise uniform
approximation of the spacetime metric is then implemented
 in Section 3, and thereby a condition for NPV propagation is derived. Section 4 is devoted
to a discussion of energy density, and the paper concludes with a
summary in Section 5.

\section{Electromagnetism in Gravitationally Affected Vacuum}

The effect of a gravitational field is captured by the metric
$\gmet$ which is a function of spacetime $x^\alpha$ and carries
the the signature $(+,-,-,-)$.\footnote{Greek indexes take the
values 0, 1, 2 and 3; Roman indexes take the values 1, 2 and 3;
$x^0 = ct$ where $c$ is the speed of light in vacuum in the
absence of all gravitational fields; whereas $x^{1,2,3}$ are the
three spatial coordinates.} In the absence of charges and
currents, electromagnetic fields obey the covariant Maxwell
equations
\begin{equation}
\label{ME1}
f_{\alpha\beta;\nu} + f_{\beta\nu;\alpha}+f_{\nu\alpha;\beta} = 0\,,\quad
h^{\alpha\beta}_{\quad;\beta} = 0\,,
\end{equation}
where
$f_{\alpha\beta}$ and $h^{\alpha\beta}$ are, respectively, the covariant and the contravariant
electromagnetic field tensors whereas the subscript $_{;\nu}$ indicates the covariant derivative with respect
to the $\nu$th spacetime coordinate.

\subsection{Noncovariant equations for vacuum}

Following common practice \c{Skrotskii}--\c{Mash},
the Maxwell equations \r{ME1} may be expressed in
noncovariant form in vacuum as
\begin{equation}
\label{ME1_noncov} f_{\alpha\beta,\nu} +
f_{\beta\nu,\alpha}+f_{\nu\alpha,\beta} = 0\,,\quad \les \le -g
\ri^{1/2}  h^{\alpha\beta} \ris_{,\beta} = 0\,,
\end{equation}
wherein $g = \mbox{det} \les g_{\alpha \beta} \ris $ and the
subscript $_{,\nu}$ denotes ordinary differentiation with respect
to  the $\nu$th spacetime coordinate. Although the generalization of the Maxwell
equations from noncovariant to covariant formulations is not completely unambiguous \c{Lightman},
we adopt the standard
generalization \r{ME1} in the absence of experimental resolution of the ambiguity.

Introduction of the electromagnetic field vectors
\begin{equation}
\left.
\begin{array}{ll}
E_\ell = f_{\ell 0}\,, \\ B_\ell = (1/2) \varepsilon_{\ell mn}f_{mn}\\
D_\ell= \le -g \ri^{1/2} h^{\ell 0}\,, \\  H_\ell=(1/2)
\varepsilon_{\ell mn} \le -g \ri^{1/2} h^{mn}
\end{array}\right\}\,,
\label{bbb}
\end{equation}
 with $\varepsilon_{\ell mn}$ being the three--dimensional
Levi--Civita symbol, allows us to state the Maxwell equations in the familiar form
\begin{equation}
\label{ME2}
\left.
\begin{array}{cc}
B_{\ell,\ell} = 0\,, & B_{\ell,0} + \varepsilon_{\ell mn} E_{m,n} = 0\\
D_{\ell,\ell} = 0\,, & -D_{\ell,0} + \varepsilon_{\ell mn} H_{m,n} = 0
\end{array}\right\}\,.
\end{equation}
 The accompanying constitutive relations of vacuum can be written for the
electromagnetic field vectors as
\begin{equation}
\label{CR2}
\left.
\begin{array}{l}
D_\ell = \gamma_{\ell m} E_m + \varepsilon_{\ell mn}\,\Gamma_m\,H_n\\[6pt]
B_\ell =    \gamma_{\ell m} H_m - \varepsilon_{\ell mn}\, \Gamma_m\,  E_n
\end{array}\right\}\,,
\end{equation}
where
\begin{equation}
\label{akh1}
\left.\begin{array}{l}
\gamma_{\ell m}
= \displaystyle{- \le -{g} \ri^{1/2} \, \frac{{g}^{\ell m}}{{g}_{00}}}\\[6pt]
\Gamma_m= \displaystyle{\frac{g_{0m}}{g_{00}}}
\end{array}\ric
\,.
\end{equation}

The most important of the foregoing equations can be expressed in
SI units as
\begin{eqnarray}
\label{eq1}
&&\curl \#E\tr + \frac{\partial}{\partial t} \#B\tr = 0\,,\\
\label{eq2}
&&\curl\#H\tr -\frac{\partial}{\partial t} \#D\tr = 0\,,\\
\label{eq3}
&&\#D\tr = \epso\,\=\gamma\tr\. \#E\tr - \frac{1}{c}\, \#\Gamma\tr\times \#H\tr\,,\\
\label{eq4}
&&\#B\tr = \muo\,\=\gamma\tr\. \#H\tr + \frac{1}{c}\,\#\Gamma\tr\times \#E\tr\,,
\end{eqnarray}
where  space $\#r$ has been separated from $t$, the scalar constants $\epso$ and $\muo$
denote the permittivity and permeability of vacuum in the absence of a gravitational
field;   $\=\gamma\tr$ is the dyadic--equivalent of $\gamma_{\ell m}$, and
 $ \#\Gamma\tr$ is the vector--equivalent of $\Gamma_m$.  These four equations are stated in the
usual style of 3--dimensional vectors and dyadics for convenience,
but the spacetime is still curved.

\subsection{Partitioning of spacetime}

Let the spacetime region of interest $\calX$ be partitioned into an appropriate number of subregions
$\calXm$, $(m= 1,2,3,\dots)$. In the $m$th subregion, the nonuniform metric $\gmet$ is written as the
sum of the uniform metric $\gmetm$ and the nonuniform residual metric $\fmetm$ as follows:
\begin{equation}
\label{eq5}
\gmet = \gmetm + \fmetm\,,\quad
\end{equation}
Notice that, whereas the curved spacetime metric $\gmet$ is
transformable into the Lorentzian metric $\eta_{\alpha\beta} =
\mbox{diag} \le 1, -1, -1, -1 \ri  $  of flat spacetime at every
point in $\calX$, in accordance with the Einstein equivalence
principle, the transformation is not universal \c{LMS05}. That is
to say, since
 $\gmet$ represents a generally curved spacetime it cannot be replaced by the flat
 spacetime metric $\eta_{\alpha\beta}$
everywhere in $\calX$.
Furthermore, there is no reason for $\gmetm$ to be transformable into $\eta_{\alpha\beta}$
at even one point in $\calXm$.

The Maxwell curl postulates read as follows in $\calXm$:
\begin{eqnarray}
\nonumber
\curl \#E\tr &=&- \les
\muo\,\gammam\. \frac{\partial}{\partial t} \#H\tr +  \frac{1}{c}\,\Gammam\times \frac{\partial}{\partial t} \#E\tr
\ris\\
\label{eq1a}
&-&
 \frac{\partial}{\partial t}
\les
\muo\,\mm\=\phi\tr\. \#H\tr +  \frac{1}{c}\,\mm\#\Phi\tr\times \#E\tr
\ris\,,
\\
\nonumber
\curl \#H\tr &=& \les
\epso\,\gammam\. \frac{\partial}{\partial t} \#E\tr -  \frac{1}{c}\,\Gammam\times \frac{\partial}{\partial t} \#H\tr
\ris\\
\label{eq2a}
&+&
\frac{\partial}{\partial t}
\les
\epso\,\mm\=\phi\tr\. \#E\tr -  \frac{1}{c}\,\mm\#\Phi\tr\times \#H\tr
\ris\,.
\end{eqnarray}
Here, $\gammam$ and $\Gammam$ are related to $\gmetm$, and $\mm\=\phi$ and $\mm\#\Phi$ to
$\fmetm$, in the same way that $\=\gamma$ and $\#\Gamma$ are related to $\gmet$.

\subsection{Piecewise Uniform Approximation}
Equations \r{eq1a} and \r{eq2a} are complicated. Therefore, for preliminary analysis, $\gmetm$ can be
selected appropriately for $\calXm$ and $\fmetm$ can be ignored. This piecewise uniform
approximation leads to the simpler equations
\begin{eqnarray}
\curl \#E\tr &=&- \les
\muo\,\mm \={\tilde\gamma}\. \frac{\partial}{\partial t} \#H\tr + \frac{1}{c}
\, \mm \#{\tilde\Gamma}\times \frac{\partial}{\partial t} \#E\tr
\ris\,,
\label{eq1b}
\\
\curl \#H\tr &=& \les
\epso\,
\mm
\={\tilde\gamma}\. \frac{\partial}{\partial t} \#E\tr -
\frac{1}{c}\,
\mm
\#{\tilde\Gamma}\times \frac{\partial}{\partial t} \#H\tr
\ris\,,
\label{eq2b}
\end{eqnarray}
for electromagnetic fields in $\calXm$.

The nature of the ``medium" implicit in \r{eq1b} and \r{eq2b} is
worth stating: This medium is spatially homogeneous and local, it
does not age, and it reacts purely instantaneously~---~just like
vacuum in the absence of a gravitational field. However, it is
bianisotropic. As $\={\tilde\gamma}$ is real symmetric, both the
permittivity and the permeability dyadics (i.e., $\epso\, \mm
\={\tilde\gamma}$ and $\muo\, \mm \={\tilde\gamma}$, respectively)
are orthorhombic and have the same eigenvectors. Furthermore, the
gyrotropic--like magnetoelectric terms on the right sides of the
two equations can be removed in the temporal--frequency domain by
a simple transform \c{LW97}, so that this medium is unirefringent
despite its anisotropy. Unless $\mm \#{\tilde\Gamma}$
 is a null vector, this medium
is not reciprocal in the Lorentz sense \c{Kr84}; despite its nonreciprocity in general, the medium satisfies the Post constraint \c{L04}. Finally, the medium
is nondissipative \cite[p. 71]{Chen}.

\section{Planewave analysis}

In this section, we turn to the propagation of plane waves in a
medium characterized by the constitutive relations \r{eq3} and
\r{eq4}, within the subregion $\calXm$ of  $\calX$. The
presuperscript $\mm$ is not generally  used henceforth in this section to
avoid cluttering up the equations. It is to be understood that all
equations in this section hold in $\calXm$ with
${\underline{\underline{\tilde\gamma}}}\equiv
\mm{\underline{\underline{\tilde\gamma}}}$ and
${\underline{\tilde\Gamma}}\equiv \mm{\underline{\tilde\Gamma}}$.

We
derive an expression for
 the time--averaged
 Poynting vector $
 \langle \#{\sf P}\rangle_t $ associated with the  plane wave with wave vector $\#k$,
and thereby establish sufficient conditions for NPV propagation,
as signalled by
\begin{equation}
\#k \. \langle \#{\sf P}\rangle_t < 0\,. \l{NPV_def}
\end{equation}

\subsection{Fourier representation}

The following three--dimensional Fourier representation of the
electromagnetic fields is appropriate for
further analysis:
\begin{eqnarray}
\#E\tr &=& \frac{1}{c}\,
\int_{-\infty}^{\infty}
\int_{-\infty}^{\infty}\int_{-\infty}^{\infty} \, \#{\sf E}\ok \,
\exp\les i(\#k\.\#r-\omega t)\ris \, d\omega
\,dk_1\,dk_2\, \label{FT1}
\\
\#H\tr &=& \frac{1}{c}\,\int_{-\infty}^{\infty}
\int_{-\infty}^{\infty}\int_{-\infty}^{\infty} \, \#{\sf H}\ok \,
\exp\les i(\#k\.\#r-\omega t)\ris \, d\omega
\,dk_1\,dk_2\,. \label{FT2}
\end{eqnarray}
Here, $i=\sqrt{-1}$,  $\#k=k_1\#{\tilde v}_1 + k_2\#{\tilde v}_2+
k_3\#{\tilde v}_3$
is the wave vector
with $k_1^2 + k_2^2 + k_3^2 =\#k\.\#k$, $\omega$ is the angular
frequency, and
\begin{eqnarray}
\#{\sf E}\ok &=&
 A_a\ok\,
 \#{\sf e}_a\ok\,
+
 A_b\ok\,
 \#{\sf e}_b\ok\, ,
\\
\#{\sf H}\ok &=&
  A_a\ok\,
 \#{\sf h}_a\ok\,
+
  A_b\ok\,
 \#{\sf h}_b\ok\, .
\end{eqnarray}
The complex--valued scalars
 $ A_{a,b}$ are unknown amplitude functions that
 can be determined from initial and boundary
conditions, and the planewave functions $\#{\sf e}_{a,b}$ and
$\#{\sf h}_{a,b}$ are as yet unspecified. Clearly, the
electromagnetic fields are thus represented in terms of an
ensemble of propagating plane waves, both propagating
(real--valued $k_3$) and evanescent (complex--valued $k_3$).

\subsection{Propagating plane waves}
Further interest being only in propagating waves, we set $k_{3}
\in \mathbb{R}$. The determination of  $\#{\sf e}_{a,b}$ and
$\#{\sf h}_{a,b}$ follows the same path as for propagation in a
simply moving medium that is isotropic dielectric--magnetic at
rest \cite[chap 8]{Chen}.

Substituting \r{FT1} and \r{FT2} in \r{eq1b} and \r{eq2b}, we find that
\begin{eqnarray}
\#p \times \#{\sf e}_{a,b}\ok &=& \omega\muo \={\tilde\gamma}
\.\#{\sf h}_{a,b}\ok\,, \l{h12}
\\
\#p \times \#{\sf h}_{a,b}\ok &=& -\omega\epso \={\tilde\gamma}
\.\#{\sf e}_{a,b}\ok\,,\l{e12}
\\
\end{eqnarray}
where
\begin{equation}
\#p = \#k -  \frac{\omega}{c} \#{\tilde\Gamma}\,.
\end{equation}
Hence,
\begin{equation}
\label{h_12}
\#{\sf h}_{a,b} \ok =
\frac{1}{\omega \muo} \={\tilde\gamma}^{-1} \. \les \#p \times \#{\sf e}_{a,b}\ok
\ris\,,
\end{equation}
while
\begin{equation}
\lec
(\#p\times\=I)\. \le {\rm adj}\, \={\tilde\gamma}\ri\.(\#p\times\=I) +
 \le\frac{\omega}{c}\ri^2 \,\={\tilde\gamma} \,\vert \={\tilde\gamma}\vert
 \ric \. \#{\sf e}_{a,b}  = \#0\,,
 \label{eee1}
 \end{equation}
 where ``adj" stands for the adjoint,  $\=I$ is the identity
 dyadic, and $\vert \={\tilde\gamma}\vert$
 is the determinant of
 $\={\tilde\gamma}$.
 As $\={\tilde\gamma}$ is symmetric, the foregoing equation
can be further simplified to
\begin{equation}
\lec \les \le\frac{\omega}{c}\ri^2 \, \vert \={\tilde\gamma}\vert
- \#p\. \={\tilde\gamma}\.\#p\ris\=I +\#p\,\#p\.
\={\tilde\gamma}\ric \. \#{\sf e}_{a,b} = \#0\,.
 \label{eee2}
 \end{equation}

Equation \r{eee2} has  nontrivial solutions if
 \begin{equation}
\Bigg \vert \les \le\frac{\omega}{c}\ri^2 \, \vert
\={\tilde\gamma}\vert - \#p\. \={\tilde\gamma}\.\#p\ris\=I
+\#p\,\#p\. \={\tilde\gamma} \; \Bigg\vert = 0\,.
\end{equation}
From this condition, the
dispersion relation
\begin{equation}
\label{dispeq}
\les\#p \. \={\tilde\gamma} \. \#p -  \le\frac{\omega}{c}\ri^2 \, \vert \={\tilde\gamma}
\vert\ris^2 = 0\,
\end{equation}
to determine $k_3$ for specific $\lec \omega,\,k_1,\,k_2\ric$
emerges. In the three--dimensional $p$--space, this relation represents
the surface of an ellipsoid \cite[Sec. 3.5.4]{BT68}.

Substituting \r{dispeq} in \r{eee2}, we obtain
\begin{equation}
\#p\,\#p\. \={\tilde\gamma}\. \#{\sf e}_{a,b} = \#0\,,
\end{equation}
whence
\begin{equation}
\#p\. \={\tilde\gamma}\. \#{\sf e}_{a,b} = 0\,.
\label{eee3}
\end{equation}
Thus, both $\#{\sf e}_a$ and  $\#{\sf e}_b$ must be  orthogonal to
$\#p\. \={\tilde\gamma}$, but neither of the two is generally
orthogonal to $\#p$. A similar exercise yields
\begin{equation}
\#p\. \={\tilde\gamma}\. \#{\sf h}_{a,b} = \#0\,,
\label{hhh3}
\end{equation}
so that both $\#{\sf h}_a$ and  $\#{\sf h}_b$ must be  orthogonal to
$\#p\. \={\tilde\gamma}$ but not necessarily to $\#p$.

The selection of $\#{\sf e}_{a,b}$ to satisfy \r{eee3} produces a trilemma, which can be explained as follows:
We can always choose two {\em unit\/} vectors $\#w$ and $\#y$ that are orthogonal to each other as well as to $\#p$. Then,
without loss of generality, \r{eee3} is satisfied by
\begin{equation}
 \#{\sf e}_a
= \frac{\={\tilde\gamma}^{-1} \. \#w}{\vert \={\tilde\gamma}^{-1} \. \#w\vert}
\,,\quad
\#{\sf e}_b
= \frac{\={\tilde\gamma}^{-1} \. (\#w + q\#y)}{\vert\={\tilde\gamma}^{-1} \. (\#w + q\#y)\vert}
\,,
\end{equation}
where $q\in \mathbb{R}$, while $\#{\sf h}_{a,b}$ can be obtained
from \r{h_12}.  The following three conditions appear reasonable  in order to fix $q$:
\begin{itemize}
\item[(i)] $\#{\sf e}_a\.\#{\sf e}_b=0$,
\item[(ii)] $\#{\sf h}_a\.\#{\sf h}_b=0$, and
\item[(iii)] $\#{\sf e}_a\times\#{\sf h}_b=\#0$ (or, equivalently, $\#{\sf e}_b\times\#{\sf h}_a=\#0$).
\end{itemize}
In general, the three conditions turn out to be mutually exclusive, i.e., only one of the three can be fulfilled.
We chose
\begin{equation}
\#w+q\#y = \#p\times \#{\sf e}_a\,
\end{equation}
in order to fulfil the third condition; thus,
\begin{equation}
 \#{\sf e}_a
= \frac{\={\tilde\gamma}^{-1} \. \#w}
{\vert\={\tilde\gamma}^{-1} \. \#w\vert}
\,,\qquad
\#{\sf e}_b = \frac{\={\tilde\gamma}^{-1}\. \le \#p \times \#{\sf e}_a
\ri}{\vert\={\tilde\gamma}^{-1}\. \le \#p \times \#{\sf e}_a
\ri\vert}
\,. \l{e_12}
\end{equation}

\subsection{Negative--phase--velocity propagation}

The rate of energy flow for a specific $\lec \omega,\#k\ric$ in
$\calXm$ is obtained by averaging the Poynting vector over one
cycle in time; thus, using  \r{h_12} and \r{e_12} we find that
\begin{eqnarray}
 \langle \#{\sf P}\rangle_t &=& \frac{1}{2 \omega \muo \vert \={\tilde\gamma}
\vert} \Big[ \vert A_a \vert^2 \#{\sf e}_a \. \={\tilde\gamma} \.
\#{\sf e}_a +  \vert A_b \vert^2 \#{\sf e}_b \. \={\tilde\gamma}
\. \#{\sf e}_b \Big] \, \={\tilde\gamma} \. \#p \,.
\label{ppp1}
\end{eqnarray}
The important result therefore emerges: the NPV condition
\r{NPV_def} is satisfied provided that
\begin{eqnarray}
 \frac{1}{ \vert \={\tilde\gamma}
\vert} \Big[ \vert A_a \vert^2 \#{\sf e}_a \. \={\tilde\gamma} \.
\#{\sf e}_a +  \vert A_b \vert^2 \#{\sf e}_b \. \={\tilde\gamma}
\. \#{\sf e}_b \Big] \, \#k \. \={\tilde\gamma} \. \#p \, < \, 0 \,.
\label{Important_result}
\end{eqnarray}

Let us elaborate on the NPV inequality \r{Important_result} for positive-- and negative--definite
 $\={\tilde\gamma}$ and indefinite  $\={\tilde\gamma}$ \c{Lut}.

\subsubsection{Positive-- and negative--definite  $\={\tilde\gamma}$}

Since $\={\tilde\gamma}$ is a 3$\times$3  dyadic, we have $ \le
\#{\sf e}_{a} \. \={\tilde\gamma} \. \#{\sf e}_{a} \ri/ \vert
\={\tilde\gamma} \vert > 0$  and $ \le \#{\sf e}_{b} \.
\={\tilde\gamma} \. \#{\sf e}_{b} \ri/ \vert \={\tilde\gamma}
\vert
> 0$  when $\={\tilde\gamma}$ is either positive--definite or
negative--definite. Thus,
 the NPV
inequality \r{Important_result} immediately reduces to
\begin{equation}
\#k \.  \={\tilde\gamma} \. \#p < 0\,. \l{npv_cond_1a}
\end{equation}
We therefore observe that NPV propagation arises for
negative--definite $\={\tilde\gamma}$ if the gyrotropic--like
magnetoelectric term $\#{\tilde\Gamma} =\#0\,$, whereas NPV
propagation is not possible for positive--definite
$\={\tilde\gamma}$ if $\#{\tilde\Gamma} =\#0$.

In order to establish the sign of  $\#k \.
\langle \#{\sf P}\rangle_t$, let us introduce the mutually orthogonal basis vectors $\#b_1, \#b_2$ and $\#b_3$,
where $\#b_3$ is parallel to $ \={\tilde\gamma} \. \#{\tilde\Gamma}$ but  $\#b_1$ and $ \#b_2$ lie in the plane
perpendicular to $ \={\tilde\gamma} \. \#{\tilde\Gamma}$. With respect to these basis vectors,
we express  $\#k$ and $ \#{\tilde\Gamma}$
as
\begin{equation}
\left.
\begin{array}{l}
\#k = \kappa_1 \#b_1 + \kappa_2 \#b_2 + \kappa_3 \#b_3\\[4pt]
 \#{\tilde\Gamma} =  \tilde{G}_1 \#b_1 + \tilde{G}_2 \#b_2 + \tilde{G}_3 \#b_3
\end{array}
\right\}.
\end{equation}
The  definiteness of $\={\tilde\gamma}$ ensures that
$\#{\tilde\Gamma}$ is not perpendicular to $ \={\tilde\gamma} \.
\#{\tilde\Gamma}$ . Thus, $\tilde{G}_3 \neq 0$ and we have
\begin{equation}
\#k = \frac{\kappa_3}{\tilde{G}_3}  \#{\tilde\Gamma} + \le
\kappa_1 - \frac{ \tilde{G}_1}{ \tilde{G}_3} \kappa_3 \ri \#b_1 +
\le \kappa_2 -  \frac{ \tilde{G}_2}{ \tilde{G}_3} \kappa_3 \ri
\#b_2\,.
\end{equation}
Equivalently, the wave vector $\#k$ may be written in the form
\begin{equation}
\#k =  \rho_1 \frac{\omega}{c} \#{\tilde\Gamma} + \rho_2 \#z\,,
\end{equation}
with  $\rho_{1,2} \in \mathbb{R}$ being scalar constants and $\#z$ being a unit vector
in the plane perpendicular to $ \={\tilde\gamma} \. \#{\tilde\Gamma}$.
It follows that
\begin{equation}
\#k \.  \={\tilde\gamma} \. \#p =\le\frac{\omega}{c}\ri^2\, \rho_1 \le \rho_1- 1
\ri\, \#{\tilde\Gamma} \. \={\tilde\gamma} \. \#{\tilde{\Gamma}}  + \rho_2^2\, \#z \.
\={\tilde\gamma} \. \#z\,.
\end{equation}
Hence, NPV propagation is a consequence of the inequality
\begin{equation}
\rho_2^2\, \#z \. \={\tilde{\gamma}} \. \#z < \le\frac{\omega}{c
}\ri^2\, {\rho_1 \le 1 - \rho_1 \ri}\, \#{\tilde{\Gamma}} \.
\={\tilde{\gamma}} \. \#{\tilde{\Gamma}}\, \l{npv_cond}
\end{equation}
being satisfied.

We emphasize that the NPV condition \r{npv_cond} applies  for an
arbitrarily oriented  wave vector $\#k$. Two particular cases are
worthy of special mention. First, if  $\#k$ lies in the plane
perpendicular to $\={\tilde\gamma}\.\#{\tilde{\Gamma}} $ (i.e.,
$\rho_1 = 0$), then NPV propagation cannot occur regardless of the
value of  $\rho_2$ or orientation of $\#z$. Second, suppose the
wave vector $\#k$ is aligned with $\#{\tilde{\Gamma}}$ (i.e.,
$\rho_2= 0$). Then the NPV inequality \r{npv_cond} is satisfied
for all $\rho_1 \in (0,1)$.

\subsubsection{Indefinite  $\={\tilde\gamma}$}

The dyadic  $\={\tilde\gamma}$  need not be  positive-- or negative--definite.
For example, the dyadic
 $\={\tilde\gamma}$ corresponding to the
 Kerr metric of a rotating black hole is indefinite in the ergosphere region \c{LMS05}.
Where  $\={\tilde\gamma}$ is indefinite, the sufficient conditions
\begin{equation}
\left.
\begin{array}{l}
\displaystyle{ \frac{1}{ \vert \={\tilde\gamma}
\vert} }\Big(  \#{\sf e}_a \. \={\tilde\gamma} \.
\#{\sf e}_a  \Big) \Big( \, \#k \. \={\tilde\gamma} \. \#p \, \Big) < \, 0 \\
\\
 \displaystyle{\frac{1}{ \vert \={\tilde\gamma}
\vert}} \Big(  \#{\sf e}_b \. \={\tilde\gamma} \.
\#{\sf e}_b  \Big) \Big( \, \#k \. \={\tilde\gamma} \. \#p \, \Big) < \, 0
\end{array}
\right\} \l{indef}
\end{equation}
 for NPV propagation emerge from \r{Important_result}.

\subsection{Summary of main results}

Let us summarize our main results in this section: For the
subregion of generally curved  spacetime $\calXm \in \calX$, with
curvature  specified by the uniform metric $\gmetm$, plane waves
propagate with phase velocity directed opposite to the direction
of the time--averaged Poynting vector provided that the inequality
\r{Important_result} holds. When $\mm\={\tilde\gamma}$ is either
positive-- or negative-- definite, the NPV inequality
\r{Important_result} reduces to the simpler relation \r{npv_cond},
whereas the sufficient conditions \r{indef} apply when
$\mm\={\tilde\gamma}$ is indefinite.

\section{Energy density}

When dealing with plane waves in linear, homogeneous materials, it is common to define the time--averaged electric and magnetic energy densities as
\begin{equation}
\left.\begin{array}{l}
\langle {\sf W}_e(\omega/c,\#k,\#r)\rangle_t = \frac{1}{4}\, {\rm Re} \les \#{\sf E}\ok\. \#{\sf D}^\ast\ok\ris \exp\les -2\, {\rm Im}\le \#k\.\#r\ri\ris
\\[6pt]
\langle {\sf W}_m(\omega/c,\#k,\#r)\rangle_t = \frac{1}{4} \,{\rm Re} \les
 \#{\sf H}^\ast\ok\. \#{\sf B}\ok\ris \exp\les -2\, {\rm Im}\le \#k\.\#r\ri\ris
\end{array}\ric\,,
\end{equation}
where the asterisk indicates the complex conjugate,
while $\#{\sf D}\ok$ and $\#{\sf B}\ok$ are defined similarly to $\#{\sf E}\ok$ in \r{FT1}. According to the Maxwell curl equations,
\begin{equation}
\left.\begin{array}{l}
\#k\times\#{\sf E}\ok = \omega\,\#{\sf B}\ok
\\[6pt]
\#k\times\#{\sf H}\ok = -\omega\,\#{\sf D}\ok
\end{array}\ric\,;
\end{equation}
therefore,
\begin{equation}
\left.\begin{array}{l} \langle {\sf
W}_e(\omega/c,\#k,\#r)\rangle_t = \frac{1}{4\omega}\,{\rm Re} \lec
\#k^\ast\. \les \#{\sf E}\ok \times \#{\sf H}^\ast\ok\ris\ric
\exp\les -2\, {\rm Im}\le \#k\.\#r\ri\ris
\\[6pt]
\langle {\sf W}_m(\omega/c,\#k,\#r)\rangle_t =
\frac{1}{4\omega}\,{\rm Re} \lec \#k\. \les \#{\sf E}\ok \times
\#{\sf H}^\ast\ok\ris\ric \exp\les -2\, {\rm Im}\le
\#k\.\#r\ri\ris
\end{array}\ric\,.
\end{equation}
The total time--averaged electromagnetic energy density $\langle {\sf W}(\omega/c,\#k,\#r)\rangle_t$ is the sum
\begin{eqnarray}
\nonumber
\langle {\sf W}(\omega/c,\#k,\#r)\rangle_t &=&
\langle {\sf W}_e(\omega/c,\#k,\#r)\rangle_t  + \langle {\sf W}_m(\omega/c,\#k,\#r)\rangle_t
\\
&=& \frac{1}{2\omega}\,{\rm Re} \le \#k\ri\. {\rm Re} \les
\#{\sf E}\ok \times \#{\sf H}^\ast\ok\ris\, \exp\les -2\, {\rm
Im}\le \#k\.\#r\ri\ris.
\end{eqnarray}
As the time--averaged Poynting vector
\begin{equation}
\langle \#{\sf P}(\omega/c,\#k,\#r)\rangle_t = \frac{1}{2}\,\les
\#{\sf E}\ok \times \#{\sf H}^\ast\ok\ris\, \exp\les -2\, {\rm
Im}\le \#k\.\#r\ri\ris,
\end{equation}
it follows that
\begin{equation}
\langle {\sf W}(\omega/c,\#k,\#r)\rangle_t = \frac{1}{\omega}\,
{\rm Re} \le \#k\ri\. \langle \#{\sf
P}(\omega/c,\#k,\#r)\rangle_t \, \exp\les -2\, {\rm Im}\le
\#k\.\#r\ri\ris.
\end{equation}
Thus, the electromagnetic energy density (as calculated in this paragraph) associated with a NPV plane wave must be negative.

Relevant to Section 3,
\begin{equation}
\left.\begin{array}{l} \#{\sf D}\ok = \epso\,\={\tilde\gamma}\.
\#{\sf E}\ok - \displaystyle{ \frac{1}{c}\,
\#{\tilde\Gamma}\times\#{\sf H}\ok}
\\ \vspace{-2mm} \\
\#{\sf B}\ok = \muo\,\={\tilde\gamma}\. \#{\sf H}\ok
+\displaystyle{ \frac{1}{c}\,\#{\tilde\Gamma}\times \#{\sf E}\ok}
\end{array}\ric\,;
\end{equation}
and the possibility of negative $\langle {\sf W}(\omega/c,\#k,\#r)\rangle_t$ for propagating
plane waves in gravitationally affected vacuum  emerges.

The possibility of a negative electromagnetic energy density   requires discussion. In the
research on isotropic, homogeneous, dielectric--magnetic NPV materials, the negative value has been noted \c{Ziol}. Equally important
is the fact that such materials have been artificially fabricated as composite materials comprising various types of
electrically small inclusions,
 and their planewave response characteristics (over limited
$\omega$--ranges) are substantially
as predicted \c{PS04}. This means implies the aforementioned procedure to compute $\langle {\sf W}(\omega/c,\#k,\#r)\rangle_t$
may not be always correct. Indeed it is not, because it applies only to nondissipative and nondispersive mediums. When
account is taken of the dissipative and the dispersive nature of the NPV
materials \c{MLW}, $\langle {\sf W}(\omega/c,\#k,\#r)\rangle_t$ does turn out to
be positive \c{Rupp}.

However, the medium in Section 3 is nondissipative and nondispersive, so that the foregoing paragraph does
not apply~---~but it does provide the basis for the following argument. Electromagnetic energy
densities for plane waves, howsoever computed, are not necessarily indicative of the true picture. This is because an
electromagnetic signal is of finite spatiotemporal extent, while
plane waves are infinitely extended over the
entire spacetime; indeed, it can be argued that a plane wave has infinite energy! Therefore, the energy density
of a signal is meaningful, but the time--averaged energy
density of a plane wave may not be. In computing the
energy density of a signal, one must consider the bandwidth
 in the $\omega\oplus\#k$ domain. Since the NPV conditions
in   Section 3 appear  unaffected by $\omega$ but not by the
direction of propagation,  NPV plane waves could appear in
gravitationally affected vacuum as part of a pulsed
electromagnetic beam (of finite cross--section) which has positive and finite energy density.

A proposal to overcome the negative value of $\langle {\sf W}(\omega/c,\#k,\#r)\rangle_t$ in
NPV materials is to fabricate them out of  {\em active\/} inclusions \c{Tret}. Whereas   {\em passive\/} inclusions are modeled
in terms of resistances, capacitances, and inductances, the modeling of active inclusions invokes amplifiers as well.
In other words, there is a source of energy to offset negative $\langle {\sf W}(\omega/c,\#k,\#r)\rangle_t$.

Reverting to \r{eq5}, we see that $\fmetm$ could be considered as
a spatiotemporally nonhomogeneous source term. The effect of this
term must be included in all energy density calculations, in
addition to the effect of the finite spatiotemporal extent of any
electromagnetic signal. In other words, the spatiotemporal
fluctuations of gravitation can act as a source term. Thus, one
must consider the total energy density, not just the
electromagnetic part of it.

Astrophysics researchers have formulated several different energy conditions   for  classical (i.e., nonquantum) GTR: all are just conjectures
lacking rigorous proofs from fundamental principles and were set up simply to prove certain theorems \c{STR01}.
Violations of these energy conditions are known \c{viol1,viol2}, and negative energy densities are invoked for
the formation of certain black holes \c{Mann} as well as for the
phenomenon of superradiant scattering of electromagnetic, gravitational, and scalar
waves \cite[Sec. 12.4]{Wald}. In fact, two astrophysicists have recently written \c{BM00}:
\begin{quote}
It is often (mistakenly) believed that every kind of matter, on scales for which we do not need
to consider its quantum features, has an energy density that is everywhere positive.
\end{quote}

This situation arises because the local energy density of a
gravitational field cannot be defined uniquely in GTR.
However, the notion of the total gravitational energy--momentum of an
isolated system~---~such as ADM energy--momentum \c{adm1}~---~is
available in an asymptotically flat spacetime; see also
\c{adm2,adm3}. But there is no guarantee that ADM total
energy should be positive. The condition of positivity of energy
can only be expected to hold if the spacetime is nonsingular and
this condition is imposed on matter distribution \c{hawkingellis}, and that for
isolated systems  \c{witten,syau}.

Under certain circumstances, many exotic solutions of general
relativity have been shown to have negative energy densities. Such
studies have exploited the use of quantum fields as possible
sources of negative energy densities \c{ford1,ford2,song}.  Unlike
classical physics, quantum physics does not restrict energy
density to have unboundedly negative values (though there are some
bounds that constrain their duration and magnitude
\c{ford1,ford2}), which then enable the quantum fields to be used
to produce macroscopic effects.

In summary, the issue of energy density remains to be carefully investigated for electromagnetic
fields in gravitationally affected vacuum, regardless of the satisfaction of the NPV condition \r{npv_cond}.
This will require numerical studies with specific spacetime metrics. A similar resolution is needed for the Casimir effect \c{Spruch,Lam}.

\section{Concluding remarks}

We have investigated the propagation of electromagnetic plane
waves in a generally curved spacetime. Sufficient  conditions
\r{Important_result}, \r{npv_cond} and \r{indef} for
negative--phase--velocity propagation are established in terms of
the spacetime metric components. We conjecture that research on new phenomenons encountered
during space exploration~---~e.g., the anomalous acceleration of Pioneer 10 currently being observed
\c{ALLLNT}~---~may benefit from NPV considerations.
The negative energy density
implications of NPV propagation require further investigation.

\vspace{10mm}

 \noindent{\bf Acknowledgement:} SS acknowledges EPSRC for support under grant
GR/S60631/01.

\end{document}